\documentclass[twocolumn,preprintnumbers,nofootinbib]{revtex4}%
\usepackage{amssymb}
\usepackage{amsmath}
\usepackage{graphicx}
\usepackage{epstopdf}
\usepackage{dcolumn}
\usepackage{bm}
\usepackage{amsfonts}
\usepackage{amsmath}
\usepackage{hyperref}%
\begin{document}
\title{Musical tonality and synchronization}
\author{Eyal Buks}
\affiliation{Andrew and Erna Viterbi Department of Electrical Engineering, Technion, Haifa 32000 Israel}
\date{\today }

\begin{abstract}
The current study is motivated by some observations of highly nonlinear
dynamical effects in biological auditory systems \cite{Goldstein_676}. We
examine the hypothesis that one of the underlying mechanisms responsible for
the observed nonlinearity is self-excited oscillation (SEO). According to this
hypothesis the detection and processing of input audio signals by biological
auditory systems is performed by coupling the input signal with an internal
element undergoing SEO. Under appropriate conditions such coupling may result
in synchronization between the input signal and the SEO. In this paper we
present some supporting evidence for this hypothesis by showing that some
well-known phenomena in musical tonality can be explained by the Hopf model of
SEO \cite{Hassard_Hopf} and the Arnold model of synchronization
\cite{Arnold_189}. Moreover, some mathematical properties of these models are
employed as guidelines for the construction of some modulations that can
be applied to a given musical composition. The construction of some intriguing
patterns of musical harmony is demonstrated by applying these modulations
to known musical pieces \cite{MuH_S}.

\end{abstract}
\maketitle





\section{Introduction}

Great deal of evidence supports the hypothesis that the dynamics of some
essential elements in the auditory system is highly nonlinear
\cite{Goldstein_676}. One of the most convincing evidence comes from
measurements of spontaneous otoacoustic emissions (OAE) that is produced by
the ear \cite{Kemp_37,Murphy_3979}. Nonlinear frequency mixing has been
observed in measurements of distortion products of OAE that are evoked using a
pair of primary tones \cite{Kujawa_142,Bian_3739}. Highly nonlinear response
of a living chinchilla's cochlea has been detected using laser velocimetric
measurements \cite{Ruggero_449}. Moreover, nonlinearity can be demonstrated in
experiments studying the perception of musical harmony, in which human
listeners are given an assignment based on a given sound that is played to
them \cite{Hartmann_3491,Langner_115,Krumhansl_334}. In particular, it was
shown that the measured perceived pitch of a sound having missing fundamental
tones reveals a nonlinear process of frequency mixing in the auditory
pathway \cite{Cartwright_5389, Cartwright_4855}.

While the above-discussed observations clearly demonstrate nonlinear dynamical
effects in the auditory system, further study is needed in order to
quantitatively explore the underlying mechanisms responsible for these
effects. It was suggested that the Hopf bifurcation model for self-excited
oscillation (SEO) can be used for the description of the auditory process of
hearing \cite{Eguiluz_5232}. In particular, it was shown that the
experimentally observed dynamic range compression \cite{Ruggero_1057} can be
related to the Hopf model. Moreover, it was suggested that the same Hopf model
can be used to describe the underlying physiological mechanisms responsible
for some universal (i.e. culture-independent) phenomena in musical tonality
\cite{Large_527,Kameoka_1460,Lee_5832}. In this description, SEO are
internally generated in parts of the human brain responsible for audio
processing of music.

In this work we further explore possible connections between the Hopf model
and musical tonality. The Hopf model provides a generic description of SEO generation,
and it allows studying the process of synchronization of SEO to externally applied
signals \cite{huygens1986pendulum,Rosenblum_401,Pikovsky_3}.  We consider a possible interpretation of
some known phenomena in musical harmony, and show that a connection between
the Hopf model and musical tonality is revealed by such an interpretation. To further validate this interpretation, some harmonic modulations are defined based on symmetrical properties of the process of synchronization. The creation of intriguing harmonies by applying these modulations to some well known musical compositions is demonstrated \cite{MuH_S}.

\section{Synchronization of SEO}

Some properties of the Hopf model that are potential  ly relevant to musical
tonality are briefly reviewed below. Consider a one-dimensional oscillator
evolving in time according to \cite{Dykman_1646}%
\begin{equation}
\dot{A}+\left(  \Gamma_{\mathrm{eff}}+i\Omega_{\mathrm{eff}}\right)
A=\xi\left(  t\right)  +\vartheta\left(  t\right)  \;, \label{A dot D}%
\end{equation}
where the complex amplitude $A$ is related to the coordinate $x\left(
t\right)  $ of the oscillator by $x=\operatorname{Re}A$ and overdot denotes a
derivative with respect to time $t$. To lowest nonvanishing order in
$\left\vert A\right\vert ^{2}$ the damping rate $\Gamma_{\mathrm{eff}}$ and
the angular resonance frequency $\Omega_{\mathrm{eff}}$ (both $\Gamma
_{\mathrm{eff}}$ and $\Omega_{\mathrm{eff}}$\ are real) are given by
$\Gamma_{\mathrm{eff}}=\Gamma_{0}+\Gamma_{2}\left\vert A\right\vert ^{2}$ and
$\Omega_{\mathrm{eff}}=\Omega_{0}+\Omega_{2}\left\vert A\right\vert ^{2}$. The
term $\xi\left(  t\right)  $ represents an externally applied force, and the
fluctuating term $\vartheta\left(  t\right)  \ $represents white noise
\cite{Risken_Fokker-Planck,Fong_023825}. In the absence of externally applied
force, i.e. when $\xi\left(  t\right)  =0$, the equation of motion
(\ref{A dot D}) describes a van der Pol oscillator \cite{Pandey_3}. Consider
the case where $\Gamma_{2}>0$, for which a supercritical Hopf bifurcation
occurs when the linear damping coefficient $\Gamma_{0}$ vanishes. Above
threshold, i.e. when $\Gamma_{0}$ becomes negative, the amplitude $\left\vert
A\right\vert $ of SEO is given by $r_{0}=\sqrt{-\Gamma_{0}/\Gamma_{2}}$ and
the angular frequency $\Omega_{\mathrm{H}}$ of SEO by $\Omega_{\mathrm{H}%
}=\Omega_{\mathrm{eff}}\left(  r_{0}\right)  $. Note that to a good
approximation the dependency of $\Omega_{\mathrm{H}}$ on $\left\vert
A\right\vert $ can be disregarded provided that $\left\vert \Omega
_{2}\right\vert \ll\Omega_{0}/r_{0}^{2}$. In what follows it will be assumed
that this dependency can be disregarded.

While the phase of SEO \cite{Rugar1989, Arcizet2006a, Forstner2012,Weig2013}
randomly diffuses in time when $\xi\left(  t\right)  =0$, phase locking
\cite{Anishchenko_117,Pandey_3,Paciorek_1723,Adler_351,Jensen_1637,DosSantos_1147}
may occur when forcing is periodically applied. Such locking results in
entrainment \cite{Hamerly_1504_04410}, i.e. synchronization
\cite{huygens1986pendulum,Rosenblum_401,Pikovsky_3} between the SEO and the
external forcing term $\xi\left(  t\right)  $ \cite{Georg_043603}.

Consider the case of a monochromatic forcing at angular frequency
$\omega_{\mathrm{d}}=\left(  1+\alpha\right)  \Omega_{\mathrm{H}}$ and
amplitude $\omega_{\mathrm{a}}$, for which $\xi$ is given by $\xi
=\omega_{\mathrm{a}}r_{0}e^{-i\omega_{\mathrm{d}}t}$. The variable $\phi$,
which is defined by $\phi=A_{\theta}+\Omega_{\mathrm{H}}t$, where
$A=\left\vert A\right\vert e^{iA_{\theta}}$, represents the phase of the
oscillator in a frame rotating at angular frequency $\Omega_{\mathrm{H}}$. Let
$2\pi\left(  Q_{n}-n\alpha\right)  $ be the value of the relative phase $\phi$
at time $t_{n}=2\pi n/\Omega_{\mathrm{H}}$, i.e. after $n$ periods of SEO,
where $n$ is integer. For the case where the change in $\phi$ over a single
mechanical period $2\pi/\Omega_{\mathrm{H}}$ is small, the evolution of
$Q_{n}$ can be described by the so-called circle map, which for the current
case is given by%
\begin{equation}
Q_{n+1}=Q_{n}+\alpha-\frac{K\sin\left(  2\pi Q_{n}\right)  }{2\pi
}\;,\label{Q RR}%
\end{equation}
where $K=2\pi\omega_{\mathrm{a}}/\Omega_{\mathrm{H}}$.

\begin{figure}
[ptb]
\begin{center}
\includegraphics[
height=2.4344in,
width=3.2396in
]%
{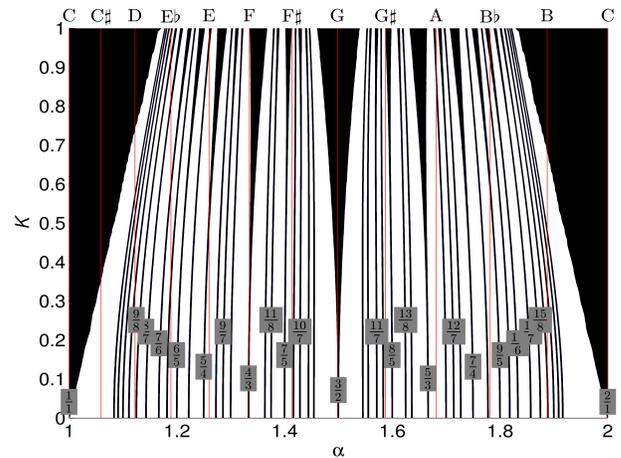}%
\caption{Arnold tongues in the plane of normalized amplitude $K$ and
	detuning $\alpha$ (only tongues corresponding to rational values $p/q$  with $q<13$ are drawn). The red lines indicate relative frequencies of the $12$ notes of the $\mu_{2}$ musical system.}%
\label{Fig_Ar}%
\end{center}
\end{figure}

The winding number $W$ is defined by $W=\lim_{n\rightarrow\infty}\left(
Q_{n+1}-Q_{1}\right)  /n$ \cite{Jensen_1637,Bak_50,Glazier_790}. For the case
of a limit cycle (i.e. phase locking), the winding number is a rational number
given by $W=p/q$, where $q$ is the period of the cycle and $p$ is the number
of sweeps through the unit interval $[0,1]$ in a cycle when the mapping
(\ref{Q RR}) is considered as modulo $1$. Regions of phase locking in the
plane that is spanned by the forcing parameters (normalized amplitude $K$ and
detuning $\alpha$) are commonly called Arnold tongues \cite{Arnold_189} (see
Fig. \ref{Fig_Ar}). Note that the graph of the function $W\left(
\alpha\right)  $ forms a structure known as a devil's staircase (which is
incomplete for $K<1$, and becomes complete when $K=1$). In the limit
$K\rightarrow0$ the width $\Delta_{\alpha}$ of the region of phase locking
corresponding to a given rational value of $W=p/q$, where $p$ and $q$ are
relatively prime integers, is given by \cite{Arnold_189}%
\begin{equation}
2\pi\Delta_{\alpha}\simeq K^{q}\;.\label{Delta alpha}%
\end{equation}
As can be seen from Eq. (\ref{Q RR}, the winding number $W$ satisfies the following symmetry relation
\begin{equation}
W\left(  \alpha\right)  =W\left(  1-\alpha\right)
\;.\label{W(alpha)=W(1-alpha)}%
\end{equation}
%

\begin{table}[tbp] \centering
	\begin{tabular}
		[c]{|c|c|c|c|c|c|}\hline
		$k$ & $d$ & $f^{\prime}$ & $\delta_{k}$ & \textbf{note} & \textbf{interval}%
		\\\hline
		$0$ & $0$ & $1$ & $0$ & $\mathrm{C}$ & $\text{perfect unison (tonic)}$\\\hline
		$7$ & $1$ & $3/2$ & $0.041$ & $\mathrm{G}$ & $\text{perfect fifth (dominant)}%
		$\\\hline
		$5$ & $-1$ & $4/3$ & $0.11$ & $\mathrm{F}$ & $\text{perfect fourth (vice
			dominant)}$\\\hline
		$9$ & $3$ & $5/3$ & $0.25$ & $\mathrm{A}$ & $\text{major sixth}$\\\hline
		$4$ & $4$ & $5/4$ & $0.32$ & $\mathrm{E}$ & $\text{major third}$\\\hline
		$3$ & $-3$ & $6/5$ & $0.40$ & $\mathrm{E}\flat$ & $\text{minor third}$\\\hline
		$8$ & $-4$ & $8/5$ & $0.42$ & $\mathrm{G}\sharp$ & $\text{minor sixth}%
		$\\\hline
		$10$ & $-2$ & $7/4$ & $0.42$ & $\mathrm{B}\flat$ & $\text{minors seventh}%
		$\\\hline
		$6$ & $6$ & $7/5$ & $0.43$ & $\mathrm{F}\sharp$ & $\text{diminished fifth,
			augmented fourth}$\\\hline
		$2$ & $2$ & $9/8$ & $0.47$ & $\mathrm{D}$ & $\text{major second}$\\\hline
		$11$ & $5$ & $15/8$ & $0.58$ & $\mathrm{B}$ & $\text{major seventh}$\\\hline
		$1$ & $-5$ & $16/15$ & $0.72$ & $\mathrm{C}\sharp$ & $\text{minor second}%
		$\\\hline
	\end{tabular}
	\caption{Sorting the musical intervals of the $\mu _{2}$ system according to their effective harmonic detunings $\delta _{k}$. The obtained ordering is almost identical to the one obtained from the measure of musical consonances (or perfection) that was suggested by Helmholtz in 1885 \cite{von1885sensations}.\label{Tab_Int}}%
\end{table}%

The spectral density of the amplitude of oscillation $x\left(  t\right)
=\operatorname{Re}A\left(  t\right)  $ has some universal properties near the
transition between the locked and the unlocked regions \cite{Strogatz_book_94}. These properties are
briefly described below for the case of the primary Arnold tongue, i.e. for
the region $\omega_{\mathrm{d}}\simeq\Omega_{\mathrm{H}}$, for which the
spectral density near the transition can be analytically evaluated. Phase
locking in this region occurs when $\left\vert i_{\mathrm{b}}\right\vert
\leq1$, where $i_{\mathrm{b}}=2\pi\alpha/K=\left(  \omega_{\mathrm{d}}%
-\Omega_{\mathrm{H}}\right)  /\omega_{\mathrm{a}}$. Outside the locking region
where $\left\vert i_{\mathrm{b}}\right\vert >1$, on the other hand, the
spectrum contains sidebands at the angular frequencies $\omega_{\mathrm{d}%
}+m\omega_{\mathrm{s}}$, where $m$ is an integer, the sideband spacing
$\omega_{\mathrm{s}}$ is given by $\omega_{\mathrm{s}}=\omega_{\mathrm{a}%
}\sqrt{i_{\mathrm{b}}^{2}-1}$, and the average frequency $\omega_{\mathrm{ave}}$ in this region is
given by $\omega_{\mathrm{ave}}=\omega_{\mathrm{d}}-\omega_{\mathrm{a}}\sqrt{i_{\mathrm{b}}^{2}-1}$.
It was shown in Ref. \cite{Buks_032202} that similar sidebands (with
similar dependency of the average frequency on the detuning) occur near the
edge of other Arnold tongues.

When $\left\vert i_{\mathrm{b}}\right\vert \gtrsim1$, i.e. just outside the
locking region, the sidebands give rise to relatively strong beats. In musical
tonality, beats commonly generate a dissonant
\cite{Burns_215,Plomp_548,von1885sensations}. On the other hand the region of
synchronization is assumed to be associated with a consonant. These musical
tonality effects are demonstrated by a media \href{https://buks.net.technion.ac.il/files/2019/10/L2UL_M.avi}{file}, which can be downloaded from \cite{MuH_S}. The audio in this file
is generated from the solution of Eq. (\ref{A dot D}) in the region
$\omega_{\mathrm{d}}\simeq\Omega_{\mathrm{H}}$ \cite{Buks_032202}. The
plot shows the spectrum for a variety of values of the normalized frequency
detuning $i_{\mathrm{b}}$, both below and above the threshold of
synchronization occurring at $\left\vert i_{\mathrm{b}}\right\vert =1$. The
dissonant nature of the sound in the region $\left\vert i_{\mathrm{b}%
}\right\vert \gtrsim1$ is apparent.

\section{Musical systems}

With a relatively weak forcing, synchronization is possible only when the
frequency ratio $\alpha$ (between forcing frequency and SEO frequency) is
sufficiently close to a rational value. A possible connection between this
property and some phenomena in musical tonality is discussed below.

A musical system with $N$ notes per octave is henceforth referred to as $N$
notes system. The frequency of the $k$'th note, where $k\in\left\{
0,1,2\cdots,N-1\right\}  $, is denoted by $F_{\mathrm{T}}f_{k}$, where
$F_{\mathrm{T}}$ is the frequency of a reference note called the tonic tone.
For an equal temperament musical tuning the relative frequencies $f_{k}$ are
equally spaced on a logaritmic scale, i.e. $f_{k}$ is given by $f_{k}=2^{k/N}
$. For this tuning method the frequency ratio between the notes $k+1$ and $k$
is a constant independent on $k$.

For a given integer $h\in\left\{  1,2,3,\cdots\right\}  $ a musical system
denoted as $\mu_{h}$ can be constructed according to the following procedure.
The rational number $\eta_{h}$ is defined to be a ratio of two adjacent primes
$\eta_{h}=p_{h}/p_{h-1}$, where $p_{h}$ denotes the $h^{\prime}$th element of
the set of prime numbers $\left\{  2,3,5,7,11,13,\cdots\right\}  $ and where
$p_{0}=1$. The integer $N_{h}$ is the number of notes per octave of the
musical system $\mu_{h}$. The notes in a given octave are labeled by a key
number $k=0,1,2\cdots,N_{h}-1$. The number $N_{h}$ is chosen such that the set
of relative frequencies $\left\{  f_{k}=2^{k/N_{h}}\right\}  $, where
$k\in\left\{  0,1,2\cdots,N_{h}-1\right\}  $, contains an element, whose key
number $k$ is denoted by $M_{h}$, for which $f_{M_{h}}=2^{M_{h}/N_{h}}%
\simeq\eta_{h}$. In general, exact match between the relative frequency
$f_{M_{h}}$ of the $k=M_{h}$ note (which can be irrational) and the rational
number $\eta_{h}$ cannot be obtained with integer finite values for both
$N_{h}$ and $M_{h}$. In practice, a mismatch of about $0.1\%$ or less is
musically acceptable. Note that the condition $f_{M_{h}}\simeq\eta_{h}$
implies that $f_{N_{h}-M_{h}}\simeq2/\eta_{h}$, i.e. the note with key number
$k=N_{h}-M_{h}$ has a relative frequency close to $2/\eta_{h}$ with respect to
the tonic. The note with key number $M_{h}$ ($N_{h}-M_{h}$) is henceforth
referred to as the dominant (vice dominant), and the following holds
$M_{h}\simeq\log_{2}\left(  \eta_{h}\right)  ^{N_{h}}$. Note that the matching
condition $f_{M_{h}}\simeq\eta_{h}=p_{h}/p_{h-1}$ can be rephrased as the
requirement that the $p_{h}$'th overtone of the tonic is close to the
$p_{h-1}$'th overtone of the dominant (in general, the frequency of the $m$'th
overtone of a note having frequency $f$ is $mf$).

For the simplest musical system $\mu_{1}$ the ratio $\eta_{1}$ is given by
$\eta_{1}=2$. The matching condition between $f_{M_{1}}=2^{M_{1}/N_{1}}$ and
$\eta_{1}$ is exactly satisfies for this case with the integers $N_{1}=1$ and
$M_{1}=1$. With these integers the musical system $\mu_{1}$ having a single note per octave is constructed.

The second system $\mu_{2}$, which is based on the rational number $\eta
_{2}=3/2$, is the common western $12$ notes per octave musical system. The integer number
$N_{2}$ is obtained from the requirement that $\log_{2}\left(  3/2\right)
^{N_{2}}$ is close to an integer. The choice of the integer $N_{2}=12$ and the
corresponding dominant key number $M_{2}=7$ and vice dominant key number
$N_{12}-M_{2}=5$ is based on the relation $\log_{2}\left(  3/2\right)
^{12}\simeq7.019\,6$. For these integers $\eta_{2}/f_{M_{2}}\simeq1.0011$,
thus the matching condition is satisfied to within an error of about $0.1\%$.

The third musical system $\mu_{3}$ is based on the rational number $\eta
_{3}=5/3$. The number of notes per octave $N_{3}=19$ and the corresponding
dominant key number $M_{3}=14$ and vice dominant key number $N_{3}-M_{3}%
=5$\ are chosen based on the relation $\log_{2}\left(  5/3\right)  ^{19}%
\simeq14.002$, and the following holds $\eta_{3}/f_{M_{3}}\simeq5/3$ to within
an error of about $0.01\%$.

Note that an alternative procedure based on the golden mean for the construction of musical scales having $N$ notes per octave has been proposed in \cite{Cartwright_51}. For each value of $N$ the musical quality of the generated scale is quantified by the so-called mean quadratic dispersion \cite{Cartwright_51}. Even though the construction procedure proposed in \cite{Cartwright_51} seems unrelated to the one described above, quite remarkably, both procedures reveal that the lowest nontrivial (i.e. larger than unity) 'good' values of $N$ are 12 and 19.

Consider a musical system $\mu$ characterized by the integers $N$ and $M$.
When $N$ and $M$ are relatively prime (i.e. the only positive integer that
divides both is $1$) it is convenient to specify to any given note having a
key number $k$ another integer called the index number $d$, which is defined
by the congruence relation $k\equiv Md$ $($mod $N)$. The inverse congruence
relation reads $d\equiv M^{-1}k$ $($mod $N)$, where the integer $M^{-1}$ is
the so-called modular multiplicative inverse. Note that $M_{2}^{-1}=7$ for the
musical system $\mu_{2}$ and $M_{3}^{-1}=15$ for the musical system $\mu_{3}$.
Both the key number $k$ and the index number $d$ specify the interval between
a given note and the tonic. However, the index number $d$ measures this
interval using the interval between the dominant and the tonic as a unit step
(e.g. the index number of the dominant note is $d=1$ and the index number of
the vice dominant note is $d=-1$). Note that the approximation $f_{k=M}%
\simeq\eta$ can be employed in order to express all $N$ relative frequencies
$f_{k}$ in terms of $\eta$ and $d$ as%
\begin{equation}
\log_{2}f_{k}\simeq\left(  \log_{2}\eta^{d}\right)  \text{ }(\text{mod }1)\;.
\label{log2(f_k)}%
\end{equation}

\section{Harmonic detuning}

While equal temperament musical tuning commonly gives rise to irrational
frequency ratios, synchronization (with a relatively weak signal) is most
efficient with rational values of ratios of small integers (i.e. rational
numbers of relatively low hierarchy in the so-called Farey tree). The level of
irrationality can be measured in a variety of ways, including the so-called
Liouville Roth irrationality exponent \cite{Liouville_133,Roth_1}. Motivated
by Eq. (\ref{Delta alpha}) for the asymptotic width of the Arnold tongues, an
alternative measure of irrationality is employed. The inaccuracy corresponding
to an approximation of a given frequency $f$ by a rational value $p/q$, where
$p$ and $q$\ are relatively prime integers, is quantified using the so-called
effective harmonic detuning $D_{p/q}\left(  f\right)  $, which is defined by
$D_{p/q}\left(  f\right)  =\left\vert f-p/q\right\vert ^{1/q}$. The infimum
harmonic detuning $D\left(  f\right)  $ is defined by $D\left(  f\right)
=\inf_{p/q\in Q}D_{p/q}\left(  f\right)  $, where $Q$ is the set of rational
numbers. While $D\left(  f\right)  =0$ for any $f\in Q$, the infimum harmonic
detuning $D\left(  f\right)  $ can take finite positive values for irrational
$f\notin Q$.

\begin{table}[h!]
	\begin{center}
		\begin{tabular}{l|c|c|c|c|c} 
			\textbf{} & \textbf{S} & \textbf{A} & \textbf{I} & \textbf{AI} & \textbf{T}\\
			\hline
			Bach C & \href{https://buks.net.technion.ac.il/files/2019/06/BachCF_U.avi}{V}  & \href{https://buks.net.technion.ac.il/files/2019/06/BachCFA_U.avi}{V} & \href{https://buks.net.technion.ac.il/files/2019/06/BachC_12To19F_U.avi}{V} & \href{https://buks.net.technion.ac.il/files/2019/06/BachCA_12To19F_U.avi}{V} & \href{https://buks.net.technion.ac.il/files/2019/06/BachCT_U.avi}{V} \\
			Mozart LD & \href{https://buks.net.technion.ac.il/files/2019/06/Mozart_U.avi}{V} & \href{https://buks.net.technion.ac.il/files/2019/06/MozartA_U.avi}{V} & \href{https://buks.net.technion.ac.il/files/2019/06/Mozart_12To19_U.avi}{V} & \href{https://buks.net.technion.ac.il/files/2019/06/MozartA_12To19_U.avi}{V} & {}\\
		\end{tabular}
	\end{center}
\caption{Musical modulations are demonstrated using the prelude in C major by Bach with the Ave Maria melody added by Charles Gounod (`Bach C'), and using Mozart Laudate Dominum (`Mozart LD'). In the columns' titles `S' stands for source, `A' for the dominant to vice-dominant modulation, `I' for the inter-system modulation from $\mu _{2}$ (12 notes per octave) to $\mu _{3}$ (19 notes per octave), 'AI' for concatenation of 'A' and 'I' modulations and `T' for the $f\rightarrow 1-f$ modulation. Red colored piano keys represent notes that are transformed to notes detached from the $\mu _{2}$ system. A variety of SoundFont files have been used for synthesizing the sound tracks. All media files can be downloaded from \cite{MuH_S}.} \label{Tab_V}
\end{table}

In the so-called just intonation tuning the irrational relative frequencies
$f_{k}=2^{k/N}$ of the equal temperament tuning are replaced by rational
numbers denoted by $f_{k}^{\prime}$. The corresponding effective harmonic
detunings are denoted by $\delta_{k}=D_{f_{k}^{\prime}}\left(  f_{k}\right)
$. Table \ref{Tab_Int} presents the calculated values of $\delta_{k}$ for the
case of the $\mu_{2}$ system sorted from small to large. The first and second
columns display the values of $k$ and $d$, respectively, the rational numbers
$f_{k}^{\prime}$ are indicated in the third column, note names are indicated
in the fifth column ($\mathrm{C}$ is chosen to be the tonic note), and the
names of the corresponding musical intervals are given in the sixth column.
This sorting suggests that the value $\delta_{k}$ can provide a useful measure
for the relative consonance level of a given musical interval (note that the
term \textit{perfect} is used only for the intervals in the top $3$ rows and
that major and minor triads can be constructed using the intervals in the top
$6$ rows). Note that a very similar ordering of musical intervals has been
obtained from an alternative sorting method based on a model of coupled neural
oscillators (see table 1 in \cite{Shapira_1429}).

As can be seen from Fig. \ref{Fig_Ar}, with finite amplitude $K$ the center of the $p/q$ Arnold tongue may shift from the point $\alpha=p/q$. Consequently, the effective harmonic detuning cannot provide a reliable measure of the harmonic importance of a given musical interval unless $K$ is sufficiently small. Note that in some cases other considerations may
affect the level of consonance. One example is the process of frequency
mixing, which may generate a tone at a frequency $f_{\mathrm{m}}$ when two
input tones at frequencies $f_{1}$ and $f_{2}$ are played together, where
$f_{\mathrm{m}}=l_{1}f_{1}+l_{2}f_{2}$, and both $l_{1}$ and $l_{2}$ are
integers. For example, for the case of the system $\mu_{2}$ having $N=12$
tones per octave, the relation $12\log_{2}\left(  1\times2^{0/12}%
+1\times2^{2/12}\right)  =13.029$ suggests that a note similar to
$\mathrm{C}\sharp$ can be generated due to nonlinearity when the notes
$\mathrm{C}$ and $\mathrm{D}$ are played together.%

A similar analysis for the case of the $\mu_{3}$ system with $N=19$ notes per
octave reveals that the lowest $6$ values of the effective harmonic detunings
are given by $\delta_{0}=D_{1}\left(  1\right)  =0$ (tonic), $\delta
_{14}=D_{5/3}\left(  f_{14}\right)  =0.052$ (dominant), $\delta_{11}%
=D_{3/2}\left(  f_{11}\right)  =0.079$, $\delta_{5}=D_{6/5}\left(
f_{5}\right)  =0.16$ (vice dominant) and $\delta_{8}=D_{4/3}\left(
f_{8}\right)  =0.18$. Note that the relatively low values of $\delta_{11}$
($f_{11}\simeq3/2$, $d=13$) and $\delta_{8}$ ($f_{8}\simeq4/3$, $d=19-13=6$) can be exploited for playing intervals in the $\mu_{3}$ system, which are
very similar to the dominant and vice dominant intervals of the $\mu_{2}$
system. Moreover, a general interval in the $\mu_{2}$ system having key number
$k$ and index number $d\equiv M_{2}^{-1}k$ $($mod $12)$, where $M_{2}^{-1}=7$,
can be imitated by playing a note having key number $k^{\prime}$ in the
$\mu_{3}$ system, where $k^{\prime}\equiv11\times d$ $($mod $19)$.

\section{Musical modulations}

The description of the synchronization process by the Hopf and Arnold models reveals some
underlying symmetries. These symmetries can be used as guidelines for the
construction of some modulations that can be applied to a given musical
piece. Some examples are discussed below. The first one (dominant to
vice-dominant modulation) is an intrasystem modulation, which is
demonstrated for the $\mu_{2}$ musical system. The second one is an
inter-system modulation, which is demonstrated by converting music from
the $\mu_{2}$ system with $12$ notes per octave into the $\mu_{3}$ system with
$19$ notes per octave. In the third example (the modulation $f\rightarrow
1-f$) notes in a given system may be converted into notes detached from any
equal temperament system. The musical modulations are demonstrated using
the prelude in C major by Bach and the Laudate Dominum by Mozart (see table \ref{Tab_V}). Media files presenting these demonstrations can be downloaded from \cite{MuH_S}.

\subsection{Dominant to vice-dominant modulation}

As was pointed out above, the relative frequencies of both the dominant and
vice-dominant notes have a relatively small effective harmonic detuning. The
so-called dominant to vice-dominant modulation is performed by replacing
the dominant frequency $\eta$ by the vice-dominant frequency $2/\eta$ in Eq.
(\ref{log2(f_k)}). The corresponding transformation of the index number $d$ is
given by $d\rightarrow d^{\prime}=A_{0}\left(  d\right)  $, where $A_{\Delta
}\left(  d\right)  =2\Delta-d$. Note that the transformation $A_{\Delta}$
mirror reflects $d$ around the point $\Delta$. Moreover, the following holds
$A_{\Delta}^{-1}\left(  d\right)  =A_{\Delta}\left(  d\right)  $ and
$A_{\Delta_{2}}\left(  d\right)  -A_{\Delta_{1}}\left(  d\right)  =2\left(
\Delta_{2}-\Delta_{1}\right)  $. Note that in the $\mu_{2}$ system with $12$
notes per octave the modulation $A_{1/2}\left(  d\right)  $ maps the
Lydian mode to the Phrygian mode, the Ionian (major) mode to the Aeolian
(minor) mode and the Mixolydian mode to the Dorian mode (and vice versa)
without changing the tonic.

\subsection{Intersystem modulation}

Replacing the dominant by another frequency having a relatively small
effective harmonic detuning can be used also for the construction of
intersystem modulations. Consider the modulation $d\rightarrow
d^{\prime}$ from a source musical system having $N$ notes per octave to a
target system having $N^{\prime}>N$ notes per octave, where $d^{\prime}=d$ for
$0\leq d\leq N/2$ and $d^{\prime}=N^{\prime}-N+d$ for $N/2<d\leq N-1$. Note
that this modulation employs only $N$ out of the $N^{\prime}>N$ notes per
octave in the target system. On a logaritmic scale, this modulation
represents a frequency multiplication by the factor $\log\eta^{\prime}%
/\log\eta$, where $\eta$ ($\eta^{\prime}$) is the relative frequency of the
dominant of the source (target) system [see Eq. (\ref{log2(f_k)})]. A
modulation from the $\mu_{2}$ source system having $12$ notes per octave
to the $\mu_{3}$ target system having $19$ notes per octave is demonstrated by the media files in table \ref{Tab_V}.

\subsection{The modulation $f\rightarrow1-f$}

The frequency modulation $f\rightarrow f^{\prime}=T_{f_{\mathrm{S}}%
}\left(  f\right)  =f_{\mathrm{S}}-f$, where $f_{\mathrm{S}}$ is the SEO
frequency, is based on the symmetry relation (\ref{W(alpha)=W(1-alpha)}). Note
that this modulation may generate notes detached from the musical system.
Harmonically satisfying results cannot be commonly obtained with a fixed
value of $f_{\mathrm{S}}$ that is kept unchanged throughout the entire musical piece. The varying value of $f_{\mathrm{S}}$ is indicated in the media files demonstrating this modulation (see table \ref{Tab_V}).

\section{Conclusions}

Dictionary definitions of the terms harmony and consonant commonly use words
such as pleasing and agreeable, whereas the words harsh and unresolved are used to define the term dissonant. The possible connection between
music tonality and SEO suggests alternative definitions that use the term
synchronizability. The synchronizability attribute can be
used to quantify the complexity of a musical piece. Catchy music has a
high level of synchronizability. On the other hand, the learning process that
makes a given composition becoming synchronizable is challenging for an
unfamiliar music having high level of complexity. Further study is needed in order to explore other implications of
synchronization on audio processing in the brain. This may help revealing the
encoding and decoding mechanisms that are employed for audio memory and audio recognition.

\section{Acknowledgments}
We thank Ivar Martin for a useful discussion.

\bibliographystyle{IEEEtran}
\bibliography{Eyal_Bib}

\end{document}